\documentclass[11pt]{article}

\usepackage[margin=1in]{geometry}
\usepackage{amsmath, amssymb, graphicx, hyperref, CJK}

\begin{document}

\begin{CJK*}{UTF8}{}

\title{Entanglement dynamics in critical random quantum Ising chain with perturbations}

\CJKfamily{gbsn}

\author{Yichen Huang (黄溢辰)\\
Institute for Quantum Information and Matter, California Institute of Technology\\
Pasadena, California 91125, USA\\
ychuang@caltech.edu}

\maketitle

\end{CJK*}

\begin{abstract}

We simulate the entanglement dynamics in a critical random quantum Ising chain with generic perturbations using the time-evolving block decimation algorithm. Starting from a product state, we observe super-logarithmic growth of entanglement entropy with time. The numerical result is consistent with the analytical prediction of Vosk and Altman using a real-space renormalization group technique.

\end{abstract}

\section{Introduction}

Entanglement, a concept of quantum information theory, has been widely used in condensed matter and statistical physics to provide insights beyond those obtained via ``conventional'' quantities. It characterizes quantum correlations in critical systems \cite{VLRK03, LRV04, CC04, CC09} and topological phases \cite{KP06, LW06, CGW10, HC15}. The scaling of entanglement \cite{ECP10} reflects physical properties (e.g., decay of correlations \cite{BH13, BH15, GH16}) and is related to the classical simulability of quantum many-body systems \cite{VC06, SWVC08, Osb12, GHLS15, Hua15}.

Many-body localization (MBL) is an active area of research studying the effects of interactions added to an Anderson insulator \cite{BAA06, Imb16, Imb16PRL, SHB+15, CHZ+16}. One characteristic feature (among others) of MBL lies in the entanglement dynamics \cite{AV15, VM16}: Starting from a product state, the entanglement entropy grows logarithmically with time. This is in contrast to the case of an Anderson insulator, in which the entanglement entropy remains bounded \cite{ANSS16}. The logarithmic growth of entanglement in MBL systems has been well established: It was observed numerically \cite{ZPP08, BPM12}, followed by theoretical explanations \cite{VA13, VA14, SPA13ent, SPA13local, HNO14}.

Besides many-body localization--delocalization transitions \cite{VHA15, PVP15}, it is also important to understand transitions between MBL phases. The critical random quantum Ising chain with generic perturbations is such an example, in which Vosk and Altman \cite{VA14} predicted super-logarithmic growth of entanglement using a real-space renormalization group (RSRG) technique. RSRG is an analytical approach to the long-range or long-time physics of random spin chains \cite{MDH79, DM80, IM05}. It is believed and only believed to be asymptotically exact at ``infinite-randomness'' quantum critical points.

The main contribution of this paper is to observe numerically the super-logarithmic growth of entanglement (which was conjectured to be a universal feature at transitions between MBL phases) using the time-evolving block decimation (TEBD) algorithm \cite{Vid04, Vid07}.

\section{Preliminaries}

We start by introducing the setup and basic definitions. Entanglement reflects a remarkable fact about the product structure of the Hilbert space for a bipartite quantum system $AB$. This Hilbert space is constructed as the tensor product of the Hilbert spaces for the two subsystems, i.e., it is spanned by product states made from (basis) vectors of $A$ and $B$. The superposition principle allows linear combinations of product states, and in general such a linear combination is not a product of any wave functions in $A$ and $B$.

The entanglement entropy of a bipartite pure state $\rho_{AB}$ is the von Neumann entropy
\begin{equation}
S(\rho_A)=-{\rm tr}(\rho_A\log_2\rho_A)
\end{equation}
of the reduced density matrix $\rho_A={\rm tr}_B\rho_{AB}$. It is the standard measure of entanglement for pure states.

The Hamiltonian of the random quantum Ising chain is
\begin{equation} \label{Ising}
H_{\rm Ising}=\sum_jJ_j\sigma_j^z\sigma_{j+1}^z+h_j\sigma_j^x,
\end{equation}
where $J_j$'s are independent and identically distributed (i.i.d.) and $h_j$'s are i.i.d. random variables. This model is non-interacting in the sense of having free-fermion representations. Its phase diagram is parametrized by
\begin{equation}
\delta=(\overline{\ln|h|}-\overline{\ln|J|})/({\rm var}\ln|h|+{\rm var}\ln|J|),
\end{equation}
which describes the competition between $J$ and $h$ terms. The spin-glass ($\delta<0$) and paramagnetic ($\delta>0$) phases are separated by an infinite-randomness critical point \cite{Fis92, Fis95}. The transition occurs in not only the ground state but also the excited eigenstates of the model \cite{HNO+13, HM14}.

The entanglement dynamics of $H_{\rm Ising}$ was studied numerically in Refs. \cite{ISL12, ZAS16}: Starting from a product state, the entanglement entropy grows double-logarithmically with time and remains bounded at and away from the critical point, respectively. This is consistent with the analytical result obtained using RSRG \cite{VA14}.

Here we study the weakly interacting model \cite{VA14, PRA+14}
\begin{equation} \label{QCG}
{\cal H}=H_{\rm Ising}+\sum_jJ'_j\sigma_j^x\sigma_{j+1}^x,
\end{equation}
where $J'_j$'s ($|J'_j|\ll|J_j|,|h_j|$) are i.i.d. random variables. The integrability-breaking $J'$ terms respect the $\mathbb Z_2$ symmetry of $H_{\rm Ising}$. Despite being irrelevant, they affect the asymptotic behavior of certain dynamical quantities. Such singular effects of integrability-breaking perturbations have been studied in other contexts; see, e.g., Refs. \cite{JHR06, JR07, HKM13}.

Let $S(t)$ denote the entanglement entropy of $|\psi(t)\rangle=e^{-i\mathcal Ht}|\psi(0)\rangle$ across a random cut in a random sample of (\ref{QCG}), where the initial state $|\psi(0)\rangle=|\uparrow\uparrow\downarrow\uparrow\cdots\downarrow\downarrow\uparrow\rangle$ is a random product state in the computational basis. Using RSRG, Vosk and Altman \cite{VA14} predicted
\begin{align}
&\langle S(t)\rangle\sim c_0\ln^{\sqrt5-1}t&{\rm for}\quad\delta=0\label{superlog}\\
&\langle S(t)\rangle\sim c_\delta\ln t&{\rm for}\quad\delta\neq0\label{log}
\end{align}
in the limit $t\to\infty$, where $\langle\cdots\rangle$ denotes averaging over randomness. Ref. \cite{VA14} did not work out the prefactor explicitly, but it is easy to see $c_0=1/8$ from the calculations there.

\section{Results}

We simulate the dynamics of (\ref{QCG}) using TEBD, which is a quite efficient method due to the slow growth of entanglement \cite{ZPP08, BPM12, VPM15}. For moderate to strong disorder, we can do very large system sizes and $t>100$ with a moderate amount of computational resources. It suffices to use only one random sample of (\ref{QCG}) and average over all cuts provided that the chain is long enough.

The simulation is set as follows. The probability density functions of the random variables $J_j,h_j,J'_j$ are $f(J_j),f(h_j),25f(25J'_j)$, respectively, where $f(x)=1/(4\sqrt{|x|})$ for $|x|\le1$ and $f(x)=0$ otherwise. This means moderate disorder and weak interaction in the sense of $|J|=|h|=25|J'|$ for typical values. The system size is $10,050$, and open boundary conditions are used. To avoid boundary effects, we average only over $10,000$ cuts in the bulk. We use a second-order Trotter decomposition with time step $0.02$, and the truncation error per step is kept below $10^{-6}$.

\begin{figure}
\centering
\includegraphics[width=0.55\linewidth]{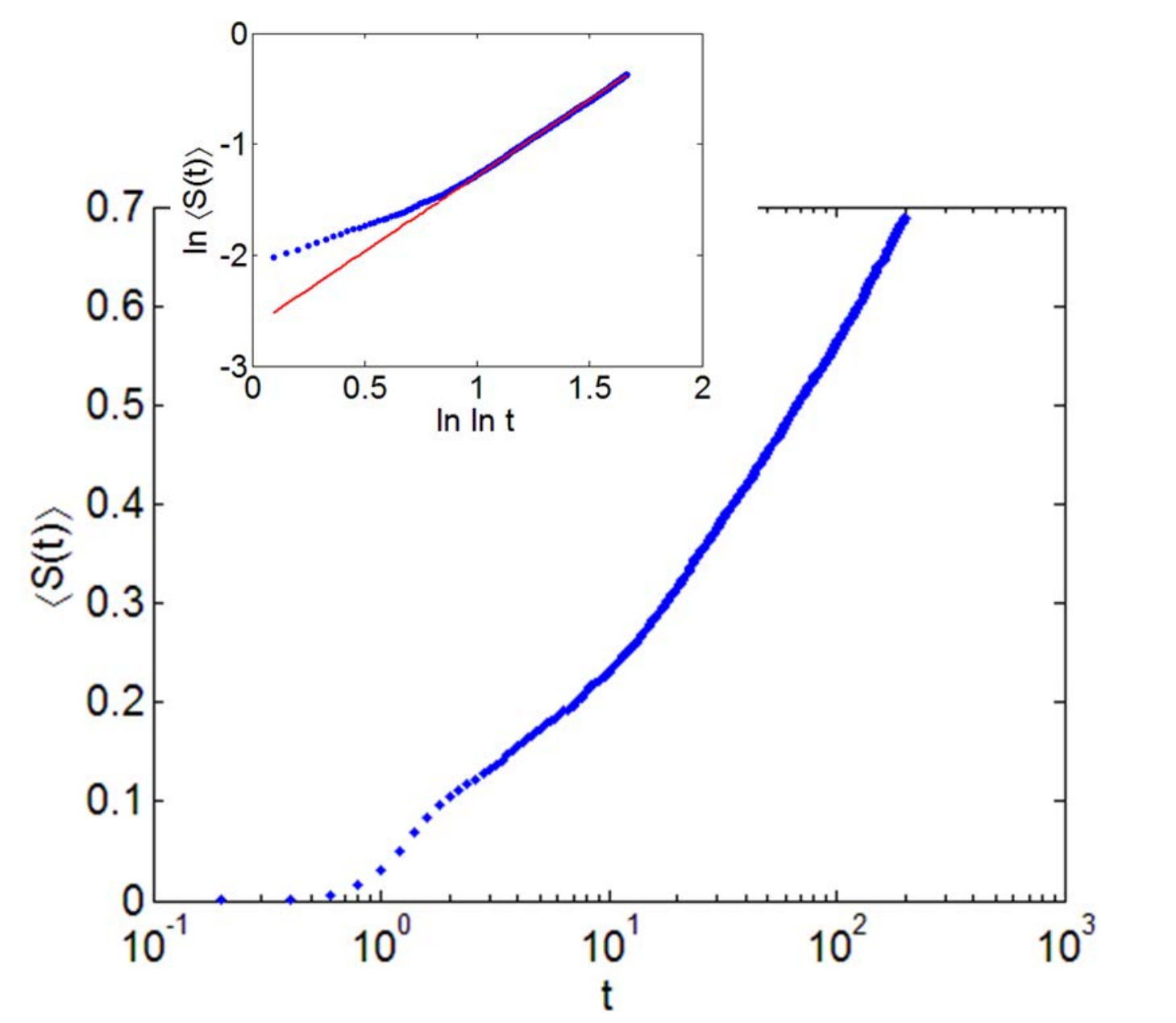}
\caption{Super-logarithmic growth of entanglement in a critical random quantum Ising chain with generic perturbations (\ref{QCG}). Inset: The same data on a loglog-log plot. The red line is a fit (\ref{fit}).}
\label{res}
\end{figure}

The simulation result is shown in Fig. \ref{res}. On a semi-log plot, the $\langle S(t)\rangle$ versus $t$ curve slightly but clearly bends upward. This is qualitatively consistent with (\ref{superlog}), which states that the entanglement entropy grows super-logarithmically with time.

It should be noted that (\ref{superlog}) only keeps the leading term in the limit $t\to\infty$. At finite $t$, a necessary condition for the subleading terms to be negligible is $\ln^{(3-\sqrt5)/2}t\gg1$ \cite{VA14}. Therefore, one cannot accurately confirm (\ref{superlog}) even if $t\approx10,000$ can be reached numerically.

Neglecting subleading terms, we demonstrate a power law relationship between $\langle S(t)\rangle$ and $\ln t$ from the data. As shown on a loglog-log plot in the inset of Fig. \ref{res}, the red line 
\begin{equation} \label{fit}
\langle S(t)\rangle=0.0709\ln^{1.36}t
\end{equation}
appears to be a good fit, which is semiquantitatively consistent with (\ref{superlog}). Noticeable differences between (\ref{superlog}) and (\ref{fit}) are expected because the subleading terms are not completely negligible. Despite this, it makes a lot of sense that the prefactors in (\ref{superlog}) and (\ref{fit}) are within a factor of $2$.

It would be desirable to confirm (\ref{log}) numerically for small $|\delta|>0$. However, a necessary condition for the effects of a finite $\delta$ to be apparent is $t\gtrsim\exp(1/|\delta|)$ \cite{VA14}. Indeed, we observe super-logarithmic growth of entanglement up to $t\approx200$ for small $|\delta|>0$ (data not shown).

\section*{Acknowledgments and Notes}

The author would like to thank Bela Bauer, Ting Cao, Joel E. Moore, Romain Vasseur, Christopher White, Pengchuan Zhang, and especially Ehud Altman for very helpful advices, comments, and suggestions.

The simulation is performed by adapting Frank Pollmann's iTEBD code\footnote{\url{http://tccm.pks.mpg.de/?page_id=871}} to random spin chains. We also recommend other public codes, e.g., the ALPS\footnote{\url{http://alps.comp-phys.org/mediawiki/index.php/Main_Page}} or ITensor\footnote{\url{http://itensor.org/}} library.

The author acknowledges funding provided by the Institute for Quantum Information and Matter, an NSF Physics Frontiers Center (NSF Grant PHY-1125565) with support of the Gordon and Betty Moore Foundation (GBMF-2644).

After completion of this work, we became aware of an arguably related paper \cite{SBYX16}, which shows a super-logarithmic ``light cone'' for out-of-time-ordered correlators (OTOC) at transitions between MBL phases. Previously, Refs. \cite{HZC17, FZSZ17, Che16, SC17, HL17, CZHF17} studied OTOC in MBL systems.

\end{document}